# Quantum transport evidence for a three-dimensional Dirac semimetal phase in Cd$_3$As$_2$

L. P. He[†], X. C. Hong[†], J. K. Dong, J. Pan, Z. Zhang, J. Zhang & S. Y. Li[*]

*State Key Laboratory of Surface Physics, Department of Physics, and Laboratory of Advanced Materials, Fudan University, Shanghai 200433, China*

[†] *These authors contribute equally to this work.*

**The material termed three-dimensional (3D) Dirac semimetal has sparked great interests recently, since it is an electronic analogue to two-dimensional graphene[1-8]. Starting from this novel phase, various topologically distinct phases may be obtained, such as topological insulator, Weyl semimetal, quantum spin Hall insulator, and topological superconductor[3]. Soon after the theoretical predictions[2,3], the angle-resolve photoemission spectroscopy and scanning tunnelling microscopy experiments gave evidences for 3D Dirac points in Na$_3$Bi and Cd$_3$As$_2$ (ref. 4-8). Here we report quantum transport properties of Cd$_3$As$_2$ single crystal in magnetic field. A sizable linear quantum magnetoresistance is observed at high temperature. With decreasing temperature, the Shubnikov-de Haas oscillations appear in both longitudinal resistance $R_{xx}$ and transverse Hall resistance $R_{xy}$. From the strong oscillatory component $\Delta R_{xx}$, the linear dependence of Landau index $n$ on $1/B$ gives an $n$-axis intercept 0.58. Our quantum transport result clearly reveals a nontrivial $\pi$ Berry's phase, thus provides strong bulk evidence for a 3D Dirac semimetal phase in Cd$_3$As$_2$. This may open new perspectives for its use in electronic devices.**

The Dirac materials whose excitations obey a relativistic Dirac-like equation have been widely studied in recent years, represented by graphene and topological insulators[8-10]. More recently, a new kind of Dirac material termed 3D Dirac semimetal has been theoretically predicted, with examples of $BiO_2$, $A_3Bi$ (A = Na, K, Rb), and $Cd_3As_2$ (ref. 1-3). The 3D Dirac semimetal possesses 3D Dirac points with linear energy dispersion in any momentum direction. The 3D Dirac point is protected by crystal symmetry, where two Weyl points overlap in momentum space[1-3]. By symmetry breakings, this 3D Dirac semimetal can be driven to topological insulator or Weyl semimetal[3]. Quantum spin Hall effect may be observed in its quantum-well structure and topological superconductivity may be achieved by carrier doping[3]. Since it is a 3D analogue to graphene, the 3D Dirac semimetal could be important for future device applications.

Following these predictions, the angle-resolved photoemission spectroscopy (ARPES) experiments were carried out on $Na_3Bi$ and $Cd_3As_2$ single crystals to probe the unique electronic structure[4-7]. Amazingly, two bulk 3D Dirac points were observed in both compounds, which locate on the opposite sides of the Brillouin zone center point $\Gamma$[4-7]. Recent scanning tunnelling microscopy (STM) measurements on $Cd_3As_2$ single crystal also support the existence of 3D Dirac points[8].

Bulk quantum transport measurement is another important tool to detect Dirac fermions. Due to their linear energy dispersion, the massless Dirac fermions should manifest a characteristic linear quantum magnetoresistance (*MR*) at the quantum limit, where all of the carriers occupy the lowest Landau level[12]. This has been previously examined for $\beta$-$Ag_2Te$ (ref. 13, 14), multilayer epitaxial graphene[15], and topological insulators[16-19]. Such a linear quantum MR was also predicted in $Cd_3As_2$, even at room temperature[3]. Dirac fermions have another distinguished feature associated with their cyclotron motions, the nontrivial π Berry's phase[20-23]. It is a geometrical phase factor, acquired when an electron circles a Dirac point. This Berry's phase can be experimentally accessed by analyzing the Shubnikov-de Haas (SdH) oscillations, which has been widely employed in the studies of graphene[22,23], graphite[24], $SrMnBi_2$ (ref. 25), topological insulators[16,18,26,27], and Rashba semiconductor BiTeI (ref. 28). Here we present the longitudinal resistance and transverse Hall resistance measurements on $Cd_3As_2$ single crystals in magnetic field. A sizable linear MR is observed near room temperature, despite that the quantum limit is not reached. We demonstrate the existence of 3D Dirac semimetal phase in $Cd_3As_2$ by observing a nontrivial π Berry's phase with a small phase shift.

Figure 1a shows the temperature dependence of the longitudinal resistivity $\rho_{xx}$ at zero magnetic field for $Cd_3As_2$ single crystal. Below 10 K the curve is very flat, giving the residual resistivity $\rho_{xx0} \approx 28.2$ μΩ cm. Figure 1b presents the Hall resistance $R_{xy}$ as a function of magnetic field from 200 K down to 1.5 K. The negative slope of $R_{xy}$ means that the dominant charge carriers in $Cd_3As_2$ are electrons, and the carrier concentration $n_e \approx 5.3 \times 10^{18}$ cm$^{-3}$ is estimated from the low-field slope. We evaluate the carrier mobility at 1.5 K as $\mu(1.5K) = 1/n_e\rho_{xx}(1.5K)e \approx 4.1 \times 10^4$ cm$^2$/Vs, where $\rho_{xx}(1.5K) = 28.2$ μΩ cm and $e = 1.6 \times 10^{-19}$ C. The low concentration and high mobility of charge carriers are well known in $Cd_3As_2$ (ref. 6-8).

In Fig. 1b, the SdH oscillations of $R_{xy}$ are already visible blow 50 K. The $R_{xx}$ shows even more pronounced oscillations, as can be seen from the longitudinal MR in Fig. 1c.

The *MR* is defined by $MR = (R_{xx}(B) - R_{xx}(0T))/R_{xx}(0T) \times 100\%$. With decreasing temperature, the *MR* increases dramatically. At 1.5 K, the *MR* oscillations can be tracked to field as low as $B \approx 2$ T, due to the high mobility of charge carriers. At 280 K, the *MR* is roughly linear, and there is no sign of saturation in high field, as high as 200% at 14.5 T. Although we observe this theoretically predicted linear *MR* (ref. 3), the quantum limit is actually not reached in our sample. There is clearly more than one Landau level occupied in our field range, as will be seen in Fig. 2a. Similar situations exist in multilayer epitaxial graphene[15], topological insulators $Bi_2Se_3$ and $Bi_2Te_3$ (ref. 16, 18, 19). To understand the physical origin of the linear *MR* without reaching the quantum limit, further theoretical study is required. Nevertheless, this large room-temperature linear *MR* is quite unusual. Its magnitude may be greatly enhanced by controlling the doping level or dimensionality of $Cd_3As_2$, which will be useful for practical applications in magnetic random access memory and magnetic sensors.

In Fig. 2a, we show the oscillatory component of $\Delta R_{xx}$ versus $1/B$ at various temperatures after subtracting a smooth background. They are periodic in $1/B$, as expected from the successive emptying of Landau levels $E_n$ when the magnetic field is increased. A single oscillation frequency $F = 58.3$ T is identified from fast Fourier transform spectra, which corresponds to $\Delta(1/B) = 0.0171$ T$^{-1}$. According to the Onsager relation $F = (\Phi_0/2\pi^2)A_k$, the cross-sectional area of the Fermi surface normal to the field is $A_F = 5.6 \times 10^{-3}$ Å$^{-2}$. By assuming a circular cross section, a very small Fermi momentum $k_F \approx 0.042$ Å$^{-1}$ is estimated. This result is in agreement with recent ARPES experiment, which shows that the Fermi surface of $Cd_3As_2$ consists two tiny ellipsoids or almost spheres[7].

The SdH oscillation amplitude can be described by the Lifshitz-Kosevich formula[28,29],

$$\Delta\rho_{xx} \propto \frac{2\pi^2 k_B T/\hbar\omega_c}{\sinh(2\pi^2 k_B T/\hbar\omega_c)} e^{-2\pi^2 k_B T_D/\hbar\omega_c} \cos 2\pi\left(\frac{F}{B} + \frac{1}{2} + \beta\right),$$

where $\omega_c$ is cyclotron frequency and $T_D$ is the Dingle temperature. $2\pi\beta$ is the Berry's phase, which will be discussed later. Figure 2b plots the temperature dependence of the normalized oscillation amplitude at $1/B = 0.0928$ T$^{-1}$, which corresponds to the 6th Landau level. The energy gap $\hbar\omega_c$ can be obtained from the thermal damping factor

$$R_T = \frac{2\pi^2 k_B T/\hbar\omega_c}{\sinh(2\pi^2 k_B T/\hbar\omega_c)}.$$

The solid line in Fig. 2b is the best fit to the data, which yields $\hbar\omega_c(n=6) \approx 24.6$ meV. For Dirac system, cyclotron frequency $\omega_c$ should follow the square root dependence on magnetic field. By employing $E_n = v_F\sqrt{2e\hbar nB}$ and $v_F = \frac{\hbar k_F}{m^*}$, rather small cyclotron effective mass $m^* \approx 0.044 m_0$ and very high $v_F \approx 1.1 \times 10^6$ m/s are obtained respectively. This value of $v_F$ is very close to the ARPES result $1.5 \times 10^6$ m/s in ref. 6. Such a large Fermi velocity may explain the unusual high mobility in $Cd_3As_2$.

Figure 3a plots together the oscillatory components of $\Delta R_{xx}$ and $\Delta R_{xy}$ at the lowest temperature $T = 1.5$ K. There are two clear features. First, the $\Delta R_{xy}$ oscillations are phase-shifted approximately by 90° with respect to the $\Delta R_{xx}$ oscillations for the low Landau levels, as expected[18]. Secondly, no Landau level splitting is observed in our field range. Figure 3b is the Landau index plot, $n$ versus $1/B$, for $\Delta R_{xx}$. We assign integer indices to the $\Delta R_{xx}$ valley positions in $1/B$ and half integer indices to the $\Delta R_{xx}$ peak positions. According to the Lifshitz-Onsager quantization rule

$A_F \frac{\hbar}{eB} = 2\pi(n + \frac{1}{2} + \beta + \delta)$, the Landau index $n$ is linearly dependent on $1/B$. $2\pi\delta$ is an

additional phase shift, resulting from the curvature of the Fermi surface in the third direction[24,28]. $\delta$ changes from 0 for a quasi-2D cylindrical Fermi surface to $\pm 1/8$ for a corrugated 3D Fermi surface[24,28]. Our data points in Fig. 3b fall into a very straight line, and the linear extrapolation gives an intercept 0.58(1). In the trivial parabolic dispersion case such as conventional metals, the Berry's phase $2\pi\beta$ should be zero. For Dirac systems with linear dispersion, there should be a nontrivial $\pi$ Berry's phase ($\beta = 1/2$). This $\pi$ Berry's phase has been clearly observed in 2D graphene[22,23] and bulk $SrMnBi_2$, in which highly anisotropic Dirac fermions reside in the 2D Bi square net[25]. For topological insulators $Bi_2Se_3$ and $Bi_2Te_3$ with gapless 2D Dirac fermions at their surfaces, the search for the $\pi$ Berry's phase was complicated by large Zeeman energy effects and bulk conduction[16,18,26,27]. The bulk Rashba semiconductor BiTeI also possesses a Dirac point and provides an alternative path to realizing the nontrivial $\pi$ Berry's phase, which was indeed experimentally detected[28]. The intercept 0.58(1) we obtain in Fig. 3b clearly reveals the $\pi$ Berry's phase, thus provides strong evidence for the existence of Dirac fermions in $Cd_3As_2$. The slight deviation from $\beta = 1/2$ indicates an additional phase shift $\delta \approx 0.08$, which may result from the 3D nature of the system[24,28].

To conclude, we have done bulk transport measurements on the proposed 3D Dirac semimetal $Cd_3As_2$ single crystals. A sizable linear quantum magnetoresistance is observed near room temperature, despite that the quantum limit is not reached. By analyzing the Shubnikov-de Haas oscillations of longitudinal resistivity in magnetic field, a nontrivial $\pi$ Berry's phase with a small phase shift is obtained. Our bulk quantum transport results unambiguously confirm the 3D Dirac semimetal phase in $Cd_3As_2$, complementary to previous ARPES and STM experiments. With its unique electronic structure, unusual high mobility, and large room-temperature linear magnetoresistance, the 3D Dirac semimetal $Cd_3As_2$ opens new avenues for future device applications.

**Methods:**

The $Cd_3As_2$ single crystals were grown from Cd flux with starting composition Cd : As = 8 : 3, as described in ref. 30. The Cd and As powders were put in an alumina crucible after sufficient grinding. The alumina crucible was placed in an iron crucible, which was then sealed in argon atmosphere. The iron crucible was heated to 825 °C and kept for 24 hours, then slowly cooled down to 425 °C at 6 °C/hour. After the alumina crucible was taken out, the excess Cd flux was removed by centrifuging in a quartz tube at 425 °C. The largest natural surface of the obtained $Cd_3As_2$ single crystals was determined as (112) plane by X-ray diffraction, with typical dimension of $2.0 \times 2.0$ mm$^2$. The quality of $Cd_3As_2$ single crystals was further checked by X-ray rocking curve, shown in the inset of Fig. 1a. The sample was cut and polished to a bar-shape, with $1.70 \times 0.78$ mm$^2$ in the (112) plane and 0.20 mm in thickness. Standard six-probe method was used for both longitudinal resistivity and transverse Hall resistance measurements. Magnetic field was applied perpendicular to the (112) plane up to 14.5 T.

**References and notes:**

1. Young, S. M., Zaheer, S., Teo, J. C., Kane, C. L., Mele, E. J. & Rappe, A. M. Dirac semimetal in three dimensions. *Phys. Rev. Lett*. **108**, 140405 (2012).

2. Wang, Z. *et al*. Dirac semimetal and topological phase transitions in $A_3$Bi (A = Na, K, Rb). *Phys. Rev. B* **85**, 195320 (2012).

3. Wang, Z., Weng, H., Wu, Q., Dai, X. & Fang, Z. Three-dimensional Dirac semimetal and quantum transport in $Cd_3As_2$. *Phys. Rev. B* **88**, 125427 (2013).

4. Liu, Z. K. *et al*. Discovery of a three-Dimensional topological Dirac semimetal, $Na_3Bi$. *Science* **343**, 864-867 (2014).

5. Xu, S. -Y. *et al*. Observation of a bulk 3D Dirac multiplet, Lifshitz transition, and nestled spin states in $Na_3Bi$. arXiv:1312.7624

6. Neupane, M. *et al*. Observation of a topological 3D Dirac semimetal phase in high-mobility $Cd_3As_2$ and related materials. arXiv:1309.7892.

7. Borisenko, S., Gibson, Q., Evtushinsky, D., Zabolotnyy, V., Buechner, B. & Cava, R. J. Experimental realization of a three-dimensional Dirac semimetal. arXiv:1309.7978.

8. Jeon, S. *et al*. Landau quantization and quasiparticle interference in the three-dimensional Dirac semimetal $Cd_3As_2$. arXiv:1403.3446.

9. Castro Neto, A. H., Guinea, F., Peres, N. M. R., Novoselov, K. S. & Geim, A. K. The electronic properties of graphene. *Rev. Mod. Phys*. **81**, 109-162 (2009).

10. Hasan, M. Z. & Kane, C. L. Topological Insulators. *Rev. Mod. Phys*. **82**, 3045-3067 (2010).

11. Qi, X. -L. & Zhang, S. -C. Topological insulators and superconductors. *Rev. Mod. Phys*. **83**, 1057 (2011).

12. Abrikosov, A. A. Quantum magnetoresistance. *Phys. Rev. B* **58**, 2788-2794 (1998).

13. Xu, R., Husmann, A., Rosenbaum, T. F., Saboungi, M. L., Enderby, J. E. & Littlewood, P. B. Large magnetoresistance in non-magnetic silver chalcogenides. *Nature* **390**, 57-60 (1997).

14. Zhang, W., Yu, R., Feng, W., Yao, Y., Weng, H., Dai, X. & Fang, Z. Topological aspect and quantum magnetoresistance of $\beta$-$Ag_2Te$. *Phys. Rev. Lett*. **106**, 156808 (2011).

15. Friedman, A. L. *et al*. Quantum linear magnetoresistance in multilayer epitaxial graphene. *Nano Lett.* **10**, 3962–3965 (2010).

16. Tang, H., Liang, D., Qiu, R. L. & Gao, X. P. Two-dimensional transport-induced linear magneto-resistance in topological insulator $Bi_2Se_3$ nanoribbons. *ACS Nano* **5**, 7510-7516 (2011).

17. He, H. *et al.* High-field linear magnetoresistance in topological insulator $Bi_2Se_3$ thin films. *Appl. Phys. Lett.* **100**, 032105 (2012).

18. Qu, D. X., Hor, Y. S., Xiong, J., Cava, R. J. & Ong, N. P. Quantum oscillations and hall anomaly of surface states in the topological insulator $Bi_2Te_3$. *Science* **329**, 821-824 (2010), and its Supporting Online Material.

19. Wang, X., Du, Y., Dou, S. & Zhang, C. Room temperature giant and linear magnetoresistance in topological insulator $Bi_2Te_3$ nanosheets. *Phys. Rev. Lett*. **108**, 266806 (2012).

20. Mikitik, G. P. & Sharlai, Y. V. Manifestation of Berry's phase in metal physics. *Phys. Rev. Lett.* **82**, 2147–2150 (1999).

21. Mikitik, G. P. & Sharlai, Y. V. Berry phase and de Haas–van Alphen effect in $LaRhIn_5$. *Phys. Rev. Lett.* **93**, 106403 (2004).

22. Novoselov, K. S *et al.* Two-dimensional gas of massless Dirac fermions in graphene. *Nature* **438**, 197–200 (2005).

23. Zhang, Y. B., Tan, Y. W., Stormer, H. L. & Kim, P. Experimental observation of the quantum Hall effect and Berry's phase in graphene. *Nature* **438**, 201-204 (2005).

24. Luk'yanchuk, I. A. & Kopelevich, Y. Dirac and normal fermions in graphite and graphene: implications of the quantum Hall effect. *Phys. Rev. Lett.* **97**, 256801 (2006).

25. Park, J. *et al*. Anisotropic Dirac fermions in a Bi square net of $SrMnBi_2$. *Phys. Rev. Lett*. **107**, 126402 (2011).

26. Analytis, J. G., McDonald, R. D., Riggs, S. C., Chu, J. H., Boebinger, G. S. & Fisher, I. R. Two-dimensional surface state in the quantum limit of a topological insulator. *Nat. Phys.* **6**, 960-964 (2010).

27. Sacépé, B. *et al.* Gate-tuned normal and superconducting transport at the surface of a topological insulator. *Nat. Commun.* **2**, 575 (2011).

28. Murakawa, H. *et al.* Detection of Berry's phase in a bulk Rashba semiconductor. *Science* **342**, 1490-1493 (2013).

29. Shoenberg, D. Magnetic oscillations in metals. *Cambridge Univ. Press*, (1984).

30. Ali, M. N., Gibson, Q., Jeon, S., Zhou, B. B., Yazdani, A. & Cava, R. J. The crystal and electronic structures of $Cd_3As_2$, the 3D electronic analogue to graphene. *Inorganic Chemistry*, doi: 10.1021/ic403163d (2014).

**Acknowledgements:** We thank D. L. Feng, J. Jiang, X. F. Jin, Y. Y. Wang, Z. Wang, Z. J. Xiang, H. Yao and Y. B. Zhang for fruitful discussions. This work is supported by the Ministry of Science and Technology of China (National Basic Research Program No. 2012CB821402), the Natural Science Foundation of China, Program for Professor of Special Appointment (Eastern Scholar) at Shanghai Institutions of Higher Learning.

**Author Contributions:** L.P.H. and X.C.H. grew the single crystals of $Cd_3As_2$. L.P.H., X.C.H., J.K.D., J.P., Z.Z. and J.Z. performed the transport measurements and analysed the data. L.P.H., J.K.D. and S.Y.L. wrote the manuscript. S.Y.L. supervised the project.

**Additional Information:** Correspondence and requests for materials should be addressed to S. Y. Li (shiyan_li@fudan.edu.cn).

**Competing financial interests:** The authors declare no competing financial interests.

**Figure 1 | Longitudinal resistivity and Hall resistance of $Cd_3As_2$.**

**a.** The longitudinal resistivity of $Cd_3As_2$ single crystal in zero magnetic field, with current in the (112) plane. The X-ray rocking curve of (224) Bragg peak is shown in the inset. The full-width at half-maximum (FWHM) is only 0.08°, indicating the high quality of the single crystals. **b.** The Hall resistance $R_{xy}$ at various temperatures. The oscillations of $R_{xy}$ are visible blow 50 K. **c.** The Shubinikov-de Haas oscillations of magnetoresistance at various temperatures, with field perpendicular to the (112) plane. The magnetoresistance is defined by $MR = (\rho_{xx}(B) - \rho_{xx}(0T))/\rho_{xx}(0T) \times 100\%$. At 280 K, the $MR$ is roughly linear without saturation, as high as 200% at $B$ = 14.5 T. At 1.5 K, the oscillations appear at field as low as 2 T, indicating the high mobility of charge carriers in $Cd_3As_2$.

**Figure 2 | Oscillatory component of longitudinal resistance.**

**a.** The oscillatory component $\Delta R_{xx}$, extracted from $R_{xx}(B)$ by subtracting a smooth background, as a function of $1/B$ at various temperatures. A single oscillation frequency $F$ = 58.3 T is identified from fast Fourier transform (FFT) spectra. **b.** The temperature dependence of the relative amplitude of SdH oscillation in $\Delta R_{xx}(B)$ for the 6th Landau level. The solid line is a fit to the Lifshitz-Kosevich formula, from which we can extract the cyclotron effective mass $m^* \approx 0.044 m_0$ and Fermi velocity $v_F \approx 1.1 \times 10^6$ m/s.

**Figure 3 | Nontrivial π Berry's phase in $Cd_3As_2$.**

**a.** The high-field oscillatory components $\Delta R_{xx}$ and $\Delta R_{xy}$ at 1.5 K. The $\Delta R_{xy}$ oscillations are phase-shifted approximately by 90° with respect to the $\Delta R_{xx}$ oscillations for the low Landau levels. No Landau level splitting is observed in our field range. **b.** Landau index *n* plotted against 1/*B*. The closed circles denote the integer index ($\Delta R_{xx}$ valley), and the open circles indicate the half integer index ($\Delta R_{xx}$ peak). The index plot can be linearly fitted, giving the intercepts 0.58(1). From the inset, the intercept of $\Delta R_{xx}$ lies between $\beta$ and $\beta$ + 1/8 ($\beta$ = 1/2), which is a strong evidence for nontrivial π Berry's phase of 3D Dirac fermions in $Cd_3As_2$.

**Figure 1**

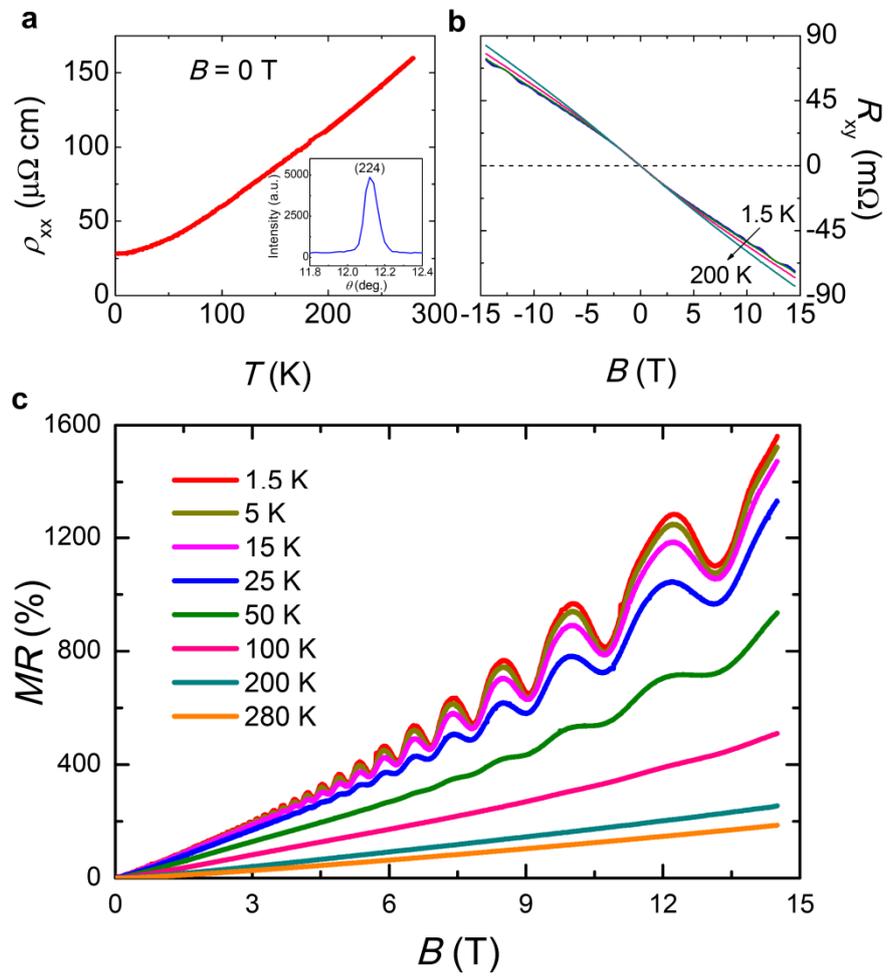

**Figure 2**

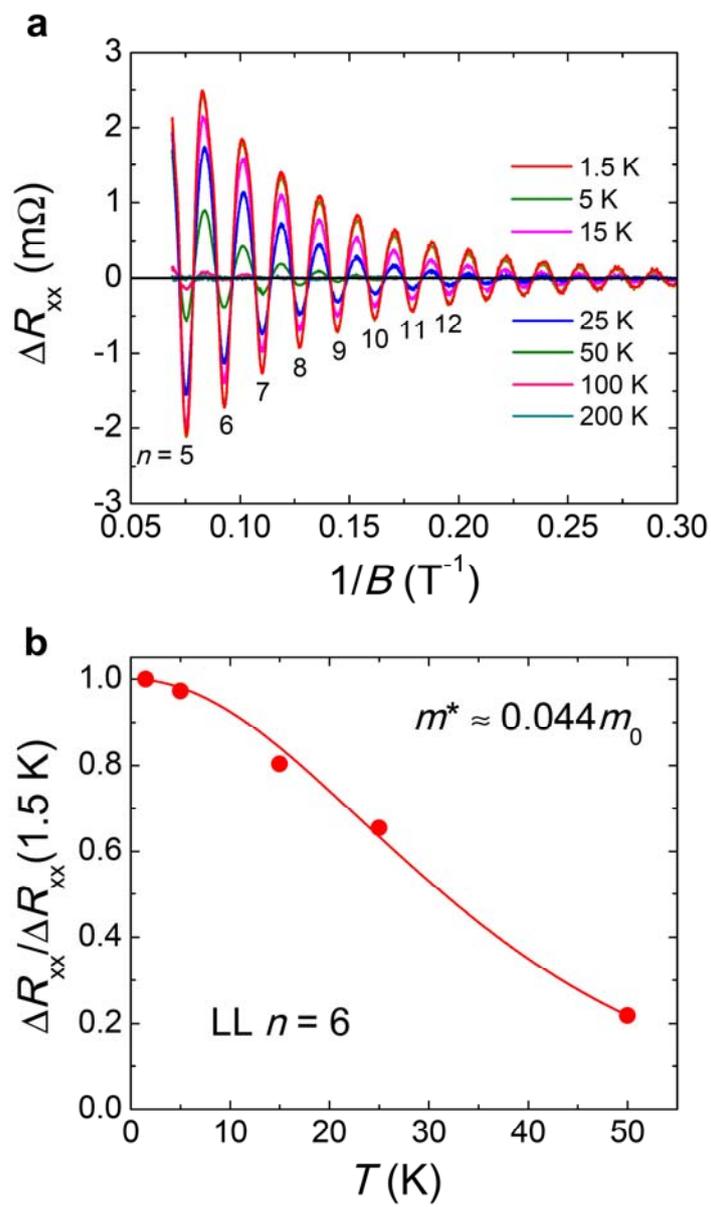

**Figure 3**

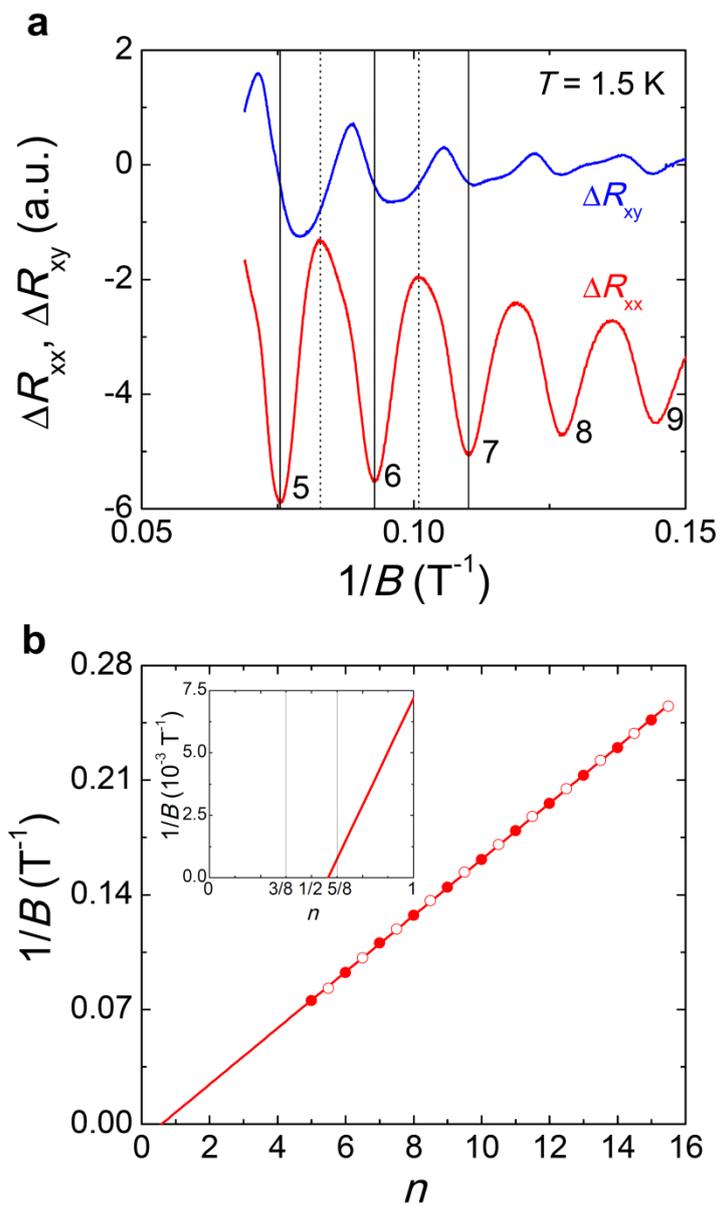